\documentclass{article}

\usepackage{arxiv}

\usepackage{graphicx}
\usepackage[utf8]{inputenc} 
\usepackage[T1]{fontenc}    
\usepackage{hyperref}       
\usepackage{url}            
\usepackage{booktabs}       
\usepackage{amsfonts}       
\usepackage{nicefrac}       
\usepackage{microtype}      
\usepackage{lipsum}		
\usepackage{multirow}
\usepackage{float}
\usepackage{url}
\usepackage{color}
\usepackage{footnote}
\makesavenoteenv{tabular}
\makesavenoteenv{table}
\usepackage[font=small,skip=0pt]{caption}
\usepackage{array}
\graphicspath{{/home/anamika/Dropbox/0_PhD/}}
\usepackage{epstopdf}
\usepackage{amsmath}
\usepackage{booktabs}
\usepackage{wrapfig,lipsum,booktabs}
\usepackage{subcaption}
\usepackage{comment}

\usepackage{booktabs}
\newcolumntype{L}{>{\centering\arraybackslash}m{1.3cm}}
\newcolumntype{C}[1]{>{\centering\let\newline\\\arraybackslash\hspace{0pt}}m{#1}}
\usepackage{mathrsfs}

\makeatletter

\title{Investigating Ortega Hypothesis in Q\&A portals: An Analysis of StackOverflow}

\date{} 					

\author{
  Anamika Chhabra \\
  Department of CSE\\
  Indian Institute of Technology Ropar\\
  India \\
  \texttt{anamika.chhabra@iitrpr.ac.in} \\
   \And
S. R. S. Iyengar \\
  Department of CSE\\
  Indian Institute of Technology Ropar\\
  India \\
  \texttt{sudarshan@iitrpr.ac.in} \\
}

\renewcommand{\headeright}{}
\renewcommand{\undertitle}{}

\begin{document}
\maketitle
\begin{abstract}
\textit{Ortega Hypothesis} considers \textit{masses}, i.e., 
a large number of average people who are not specially qualified as being instrumental in any system's progress. This hypothesis has been reasonably examined in the scientific domain where it has been supported by a few works while refuted by many others, resulting in no clear consensus. While the hypothesis has only been explored in the scientific domain so far, it has hardly been examined in other fields. Given the large-scale collaboration facilitated by the modern Q\&A portals where a crowd with a diverse skill-set contributes, an investigation of this hypothesis becomes necessary for informed policy-making. In this work, we investigate the research question inspired from Ortega Hypothesis in StackOverflow where we examine the contribution made by masses and check whether 
the system may continue to function well even in their absence. 
The results point towards the importance of masses in Q\&A portals for the little but useful contribution that they provide. The insights obtained from the study may help in devising informed incentivization policies enabling a better utilization of the potential of the users.
\end{abstract}

\keywords{Ortega Hypothesis\and Newton Hypothesis\and StackOverflow\and Masses\and Incentivization Policies.}

%
%
%

\section{Introduction}


There are two different schools of thought when it comes to attributing value to people with different contribution levels in any system. The first one, known as \textit{Ortega Hypothesis}, regards a mass of medium level contributors as being instrumental in the system's overall functioning~\cite{ortega1930revolt,cole1972ortega}. The hypothesis is attributed to \textit{Ortega Y. Gasset}, who in his book `The Revolt of the Masses'~\cite{ortega1930revolt} highlighted the importance of \textit{masses} in any field. Ortega defines masses as average people who are not `specially qualified'. The importance of masses was also supported by Florey~\cite{crowther1968science} who asserted that there is nothing called a \textit{breakthrough} in science, rather coming up with an excellent piece of work requires small inputs by a large number of people. Many other studies also support the capabilities of masses for doing good science~\cite{travis2008science}. On the other hand, an opposing hypothesis called \textit{Newton Hypothesis} supports the view that only a bunch of top-level contributors, i.e., \textit{elites} are sufficient for making progress in any field and the remaining mass of medium and low-level contributors may be safely discarded~\cite{cole1972ortega}. These opposing hypotheses have been examined in the scientific domain using bibliometric analyses, although without any clear consensus~\cite{cole1972ortega,macroberts1987testing,cole1987testing,hoerman1995secondary}. The research community seems divided with respect to their acceptance or rejection of Ortega hypothesis~\cite{macroberts1987testing,turner1976another,bornmann2010scientific}.


The importance of medium (or low) level contributors has hardly been explored in fields other than the scientific domain. Sometimes referred to as \textit{core} and \textit{peripheral} users, masses and elites have been examined to a limited extent in peer-production communities~\cite{arazy2014wikipedia,borgatti2000models}. However, not much has been done towards examining their relative importance with respect to the functioning of the system. The existing state-of-the-art in this direction reveals our inability to understand and acknowledge users' importance with respect to their contribution. This has implications for incorrect and sub-optimal policies on collaborative portals. Further, with the recent development of the online collaboration tools that rely on the contribution by a large number of users in the online crowd, it becomes important to appropriately examine and reward the contribution made by users at different levels. The users in online crowdsourced environments come with a diverse set of expertise, background and motives, leading to a highly unequal contribution~\cite{nielsen2006participation,stewart2010crowdsourcing,haklay2016participation,jakob2006inequality}. This inequality of contribution has been observed in diverse settings such as Wikipedia~\cite{ortega2008inequality,arazy2010determinants}, Wikis~\cite{serranio_inequality_wiki}, Usenet Newsgroups~\cite{whittaker2003dynamics}, Stackoverflow~\cite{wang2013empirical,yang2014sparrows}, Internet, blogs and Amazon~\cite{jakob2006inequality}. While such inequality has been observed to automatically emerge in collaborative settings~\cite{muchnik2013origins,yang2010motivations,kittur2010beyond}, its implications have rarely been investigated. In this work, we focus on Q\&A portals by analyzing the most popular portal for programming related questions, i.e., StackOverflow. It is the oldest in the StackExchange network that contains more than 156 Q\&A portals each covering a particular topic. 
StackOverflow contains more than 16 million questions, 25 million answers and 9.2 million users~\cite{SE2019}
On average, around 9.9 million users visit the website everyday and about 7200 questions are asked each day. Given the popularity and the availability of large-scale data of this website, we found it suitable as well as important for an analysis of masses' contribution on Q\&A portals.


%
%
%
We divide the users of StackOverflow into different categories based on the extent of their contribution which we measure in terms of the questions and answers posted by them. We define masses to be the users with a limited engagement- as compared to 
a bunch of highly active users, i.e., elites - but constituting the largest pool of users in the portal. If Newton hypothesis is to be true, then the contribution by elites will be self-sufficient for the portal, with the contribution by masses being trivial. Further, if Ortega hypothesis is to be true, then masses should be providing useful and indispensable contribution to the portal. 
We compare the overall contribution made by masses with that of the elite bunch across quantitative and qualitative dimensions. Our analysis shows that the elite users are essential for a Q\&A portal as they are the primary source of answers. However, of particular interest is the observation that these elite users are considerably dependent on masses for the questions that they ask. We also use \textit{HITS algorithm}~\cite{kleinberg1999authoritative} to compare the questioning and answering capability of masses and elites in the network of `who-answers-whom' and find a lack of questioning capability in elites as compared to masses. Knowing that in a Q\&A portal, questions are responsible for triggering answers into the system, the users asking questions naturally become important. The results reflect the importance of masses and point towards Ortega hypothesis to be valid in Q\&A portals, while at the same time not ruling out the importance of elite users. In particular, the results suggest that a Q\&A portal whose policies are designed \textit{only} considering the Newton Hypothesis to be true might not be able to achieve its intended purpose.



The study finds relevance for designing appropriate interfaces and policies for motivating the less-active users in Q\&A portals. 
Most of these portals employ various incentivization procedures which award points, badges or special privileges to users based on their contribution~\cite{anderson2013steering,easley2016incentives}. These procedures are known to affect users' engagement level and style~\cite{hamari2014does,antin2011badges,hamari2017badges}. Therefore, they need to be carefully designed. Currently, they are mostly devised keeping only the small bunch of elite users into consideration, while not giving much weight to the majority of mass users~\cite{easley2016incentives}. 
For instance, Amazon and Yahoo Answers employ a \textit{relative} reward mechanism by acknowledging the top contributors. Such policies discourage a huge cohort of users who may not be able to reach these standards with their little piece of contribution. 
Devising educated strategies thus may help in retaining useful information providers in these portals. 

\section{Related Work}
\subsection{Scientific Domain}


In 1930, in his book, Ortega  asserted the importance of the `average scientist' in advancing the sciences~\cite{ortega1930revolt}. He emphasized that the average scientists provide the support and help to the top-scientists, however, their contribution remains unacknowledged. He stated that these people are either mediocre or even less than mediocre, for example, people working in labs, conducting experiments, hence helping the great discoverers. The book however, highlights the importance of average people not only in scientific domain, rather in other areas such as history and politics as well. As stated earlier, Ortega hypothesis has so far been examined only in scientific domain, although without any general consensus. 
The first work in this direction was conducted by Cole and Cole~\cite{cole1972ortega} through their citation analysis where they argued that only a few scientists contribute to the scientific progress. The authors checked the best work of a group of 84 physicists of a university and analyzed the work that they had cited to come up with their own discoveries. They observed that these works were mostly citing the highly cited work and hence disproved Ortega Hypothesis.  
The authors' analysis had this assumption that citation count is an indicator of influence of a work. However, Macroberts and Macroberts~\cite{macroberts1987testing} later questioned their work arguing that scientists do not always cite the work that they use in their research. They concluded that a proper test of the Ortega hypothesis is yet to be done. Another work by Hoerman et al. \cite{hoerman1995secondary} studied the referencing behavior of researchers and examined many irregularities in the citing practices of researchers. The latest work in this direction was carried out in 2010 by Bornmann et al.~\cite{bornmann2010scientific} which was also based on the citation analysis. The authors analyzed the papers from Scopus and Web of Science published in 2003 in different subject areas and found that highly-cited papers were mostly citing highly-cited papers, hence supporting Newton Hypothesis, and refuting Ortega hypothesis in scientific domain.

\subsection{StackOverflow}


A few studies have been carried out on StackOverflow examining the inequality of contribution made by users. 
These studies show that StackOverflow comprises of a large number of users making a tiny amount of contribution. In particular, based on the analysis of a dataset comprising of 63,863 questions, Wang et al.~\cite{wang2013empirical} observed that only 7.8\% of users answered more than five questions and only 1.6\% of the users asked more than five questions. A good proportion of users asked only one question or provided only one answer. A related study on StackExchange examining the contribution of users based on their reputation was conducted by  Movshovitz-Attias et al.~\cite{movshovitz2013analysis} where they found that  
%
most of the answers to the website come from the high reputation users.
The authors' division of users was based on the reputation that they gain through the incentivization system of StackExchange. 

While the inequality of contribution has been observed in StackOverflow, the worth of users with respect to their contribution has not been investigated.
%



\section{Data Set}

\begin{figure}[!htb]
\centering
  \begin{tabular}{@{}c@{\hskip1pt}c@{}}
	\includegraphics[width=0.35\textwidth]{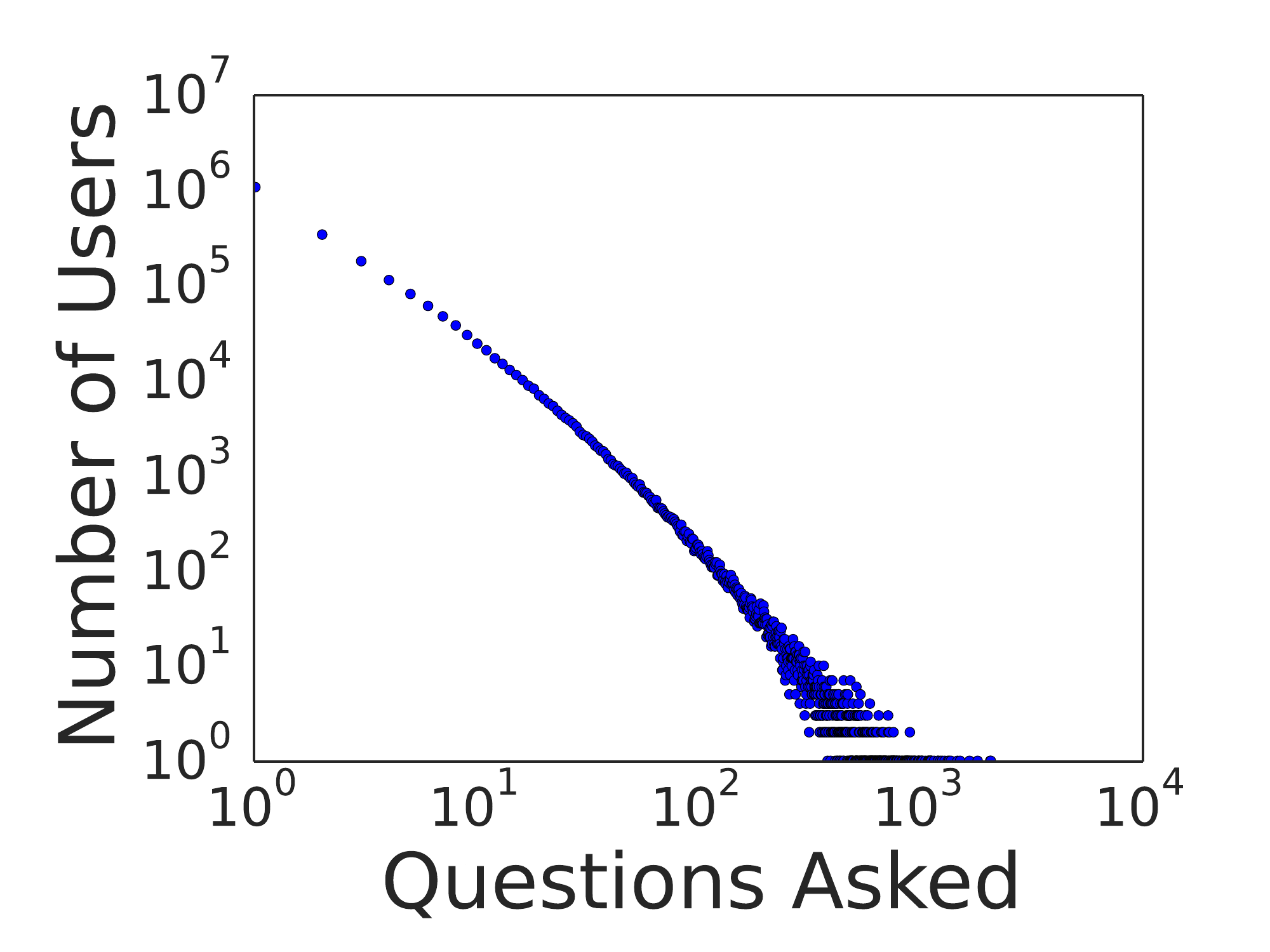} &
	\includegraphics[width=0.35\textwidth]{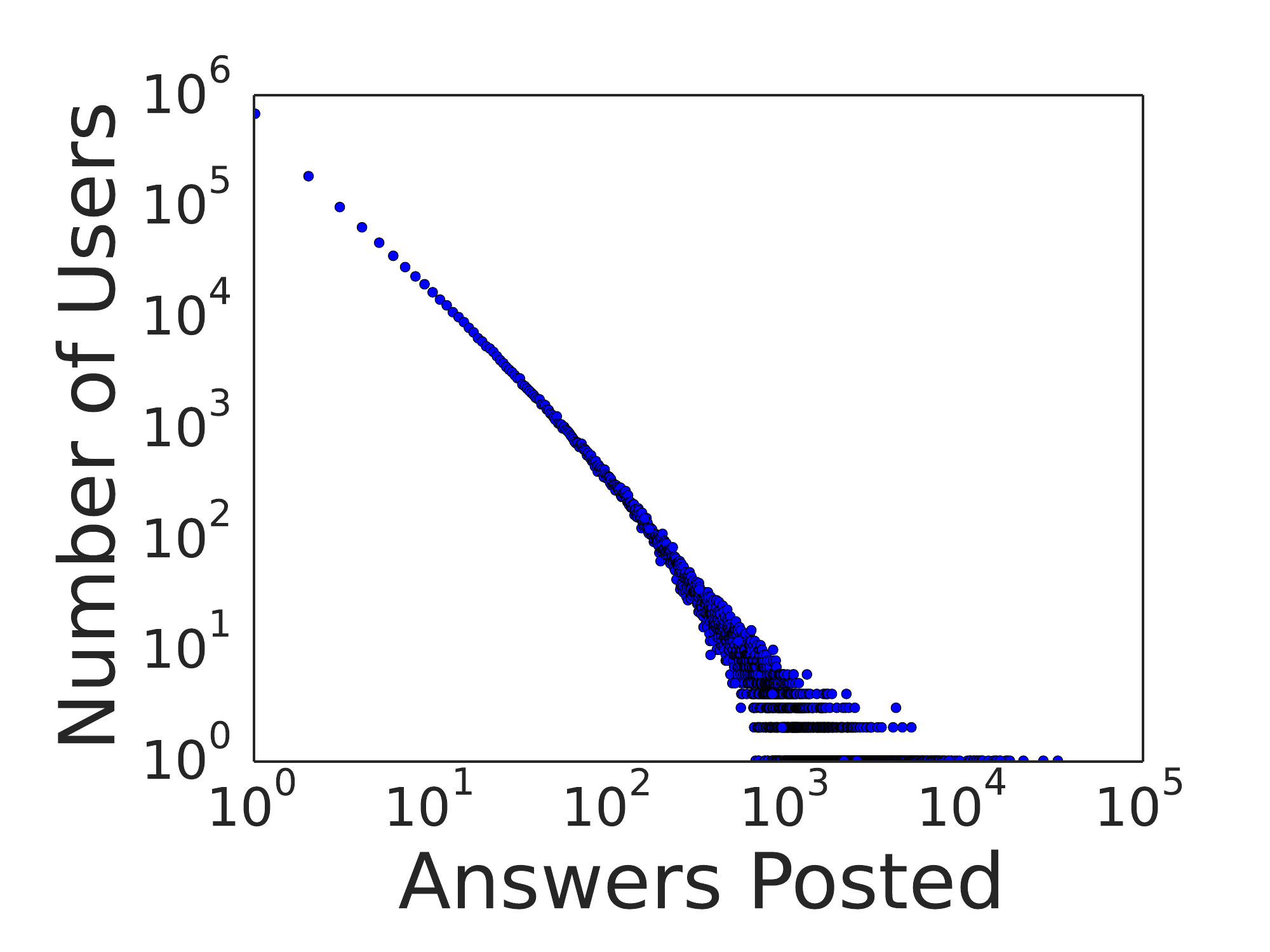}\\
	\textbf{(a)}  & \textbf{(b)}
  \end{tabular}
\caption{Distribution of users as per the number of (a) questions asked and (b) answers posted. The distributions exhibited a power law behavior, indicating highly unequal contribution.} \label{fig:qa_dist}
\end{figure}

The dataset of StackOverflow was downloaded in August 2016 from the publicly available archive~\cite{SE2016dump}
amounting to 141 GB of data. It consists of the details of all the question threads along with the information about answers, users, tags, links, votes, badges, suggest-edits, deleted posts, timestamps etc in XML format. 

To check the disparity in users' contribution, we computed the distribution of users as per the number of questions and answers posted by them respectively. Figure~\ref{fig:qa_dist} shows the log-log plot of the distribution of users as per (a) the number of questions asked and (b) the number of answers posted. A large fraction of users was found to be posting very few questions and answers. Moreover, the fraction of users who were contributing a lot was very less. The plots confirm the contribution inequality in StackOverflow. Further, both the distributions followed a power law with exponent $\gamma$ = 4.2 for questioning and $\gamma$ = 3.0 for answering. 
High values of power law exponents establish an exponential drop in the proportion of users making a large amount of contribution. In particular, 48.71\% of the users asked only one question and 47.72\% of the users posted only one answer. This observation corroborates the finding of Wang et al.~\cite{wang2013empirical} about a large number of users making a very small amount of contribution on StackOverflow. 
Additionally, the maximum number of questions asked by a single user was 2,061 and the number of answers posted by a single user was as high as 33,135. It is important to remark that in this analysis, we examine only those users who posted at least one question or answer.\footnote{Out of a total of 6,455,132 users, 56.37\% of them did not add even a single question or answer. They registered on the website and mostly remained passive consumers of content.}
Also, we took care of not considering the contribution made by the \textit{Community User} which is an automated process that performs multiple background activities as well as owns community questions and answers~\cite{SE2019community}.

\section{Analysis and Results}
This Section describes how we divided the users of StackOverflow into categories for comparison and then reports the observations made on the contribution by these categories.
\subsection{Percentile Classes}

The literature points to the existence of specialism in activity selection by users, where they exhibit roles on crowdsourced websites, thereby making most of their contribution in one of the activities~\cite{nam2009questions,chhabra2015presence,chhabra2015characteristic,chhabra2015skillset}. On Q\&A portals particularly, users are observed to play the role of \textit{askers} or \textit{solvers}~\cite{nam2009questions}. Therefore, it may be possible that the users who are posting very few answers may be contributing heavily in questioning and vice versa. Further, a majority of the existing studies have been focusing only on users who provide answers. This is indicated by an active research in the direction of identifying or predicting expert answerers~\cite{jurczyk2007discovering,zhang2007expertise,bouguessa2008identifying,pal2012evolution,riahi2012finding}. This is reasonable enough, knowing that these users are the resource persons and provide solutions to other users' problems.
Nevertheless, in a Q\&A portal, the importance of users asking questions may not be ignored, as in the absence of these questions, the expert answerers will also not be able to contribute. Therefore, from a systemic perspective, questions make an important contribution in a Q\&A portal. Since our analysis aims to examine users' overall contribution in primary activities, we track each user's contribution in both questioning and answering. We specifically do not consider the activities such as commenting and editing, as these are side activities on StakOverflow and a minimum reputation score is required before a user can contribute in them. Further, we use votes to judge the quality of users' contribution. 

To compute each user's overall interaction with the portal, we add up their total contribution in questioning and answering and call this measure \textit{qa-count} for that user. Clearly, this measure gives equal weightage to both questions and answers. This is a deliberate decision as it primarily captures the extent of users' interaction with the portal and will help in the stratification of users. Later, while investigating users at different levels, we consider their contribution in questioning and answering separately (Subsection~\ref{qa_activity}).

The qa-count was found to range from 1 to 33174. Given this wide spectrum as well as the interdependency between questions and answers, we followed the method adopted by Bornmann et al.~\cite{bornmann2010scientific} and Green et al.~\cite{green1981test} to divide the users into percentile classes. The authors in their bibliometric analysis, divided the researchers into six percentile classes based on their citation count. 
%
%
Similarly, we sorted the users based on their qa-count and divided them into percentile classes namely $lt_{50}$, $bt_{50-75}$, $bt_{75-90}$, $bt_{90-95}$, $bt_{95-99}$ and $gt_{99}$ having users with less than 50, 50 to 75, 75 to 90, 90 to 95, 95 to 99 and greater than 99 percentile values of qa-count respectively. Clearly, highly-active top 1\% of the users belong to the class $gt_{99}$. 
\begin{table}[!htp]
\scriptsize
\centering
\begin{tabular}{>{\bfseries}p{2.3cm}|p{2.3cm}|p{2.3cm}}
\hline
Category & \textbf{Percentile \newline Class} & \textbf{Range of \newline qa-count} 	\\

\hline
\multirow{4}{*}{LC} & lt50 & 1-2 \\
& bt\_50\_75 & 2-5\\
& bt\_75\_90 & 5-17 \\
& bt\_90\_95 & 17-37\\
\hline
MC & bt\_95\_99 & 37-151 \\
\hline
TC & gt\_99 & 151-33135 \\

\hline
\end{tabular}
\captionof{table}{Range of qa-count for the users of percentile classes. (qa-count is the total number of questions and answers posted by the users.)}
\label{p_classes}
\end{table}

Table~\ref{p_classes} shows the range of qa-count observed for the users belonging to each class. Interestingly, the sum of questions and answers posted by the users in class $lt_{50}$ was either 1 or 2 only, whereas this class contains 50\% of the total number of users. Similarly, the Table shows that more than 95\%
of the users (from $lt_{50}$ to $bt_{90-95}$) posted less than a total of 37 questions and answers. Further, only 4\% of the users (class $bt_{95-99}$) posted a total of 37 to 151 questions and answers. Finally, the users of $gt_{99}$ class reported a very high contribution ranging from a total of 151 to 33,135 questions and answers. In the subsequent text, we will be reporting results obtained for each of these classes. Additionally, to be able to make a higher-level comparison in the direction of Ortega Hypothesis, we also provide an alternate categorization, which we call \textit{LC-MC-TC categorization}. Such a categorization was used by Cole and Cole~\cite{cole1972ortega} for their bibliometric analysis on scientists' worth from different strata. We refer to users belonging to the classes from $lt_{50}$ to $bt_{90-95}$ (comprising of 95\% of the users) as `Low Contributors (LC)', those belonging to $bt_{95-99}$ as `Medium-level Contributors (MC)', while the last class, i.e., $gt_{99}$ as `Top Contributors (TC)'. 
This categorization on top of the percentile classes will help in obtaining a broader perspective apart from a contribution comparison across a finer-grained spectrum. For easy visualization, we would be showing comparison among the categories LC, MC and TC, especially focusing on the two extreme categories, i.e., LC and TC that represent users from masses and elites respectively.

%
%
%


\subsection{Questioning and Answering Activity of the Classes}\label{qa_activity}
We now compare the questioning and answering contribution of the users belonging to each class along with the qualitative parameters such as votes and \textit{favorite counts} obtained on their content. We also examine the extent of dependency prevalent among the users of these classes.

\subsubsection{Quantitative Analysis}\label{quant}

%
%
%



%

\begin{figure}
  \begin{minipage}[b]{0.45\textwidth}
    \centering
\includegraphics[scale = 0.26]{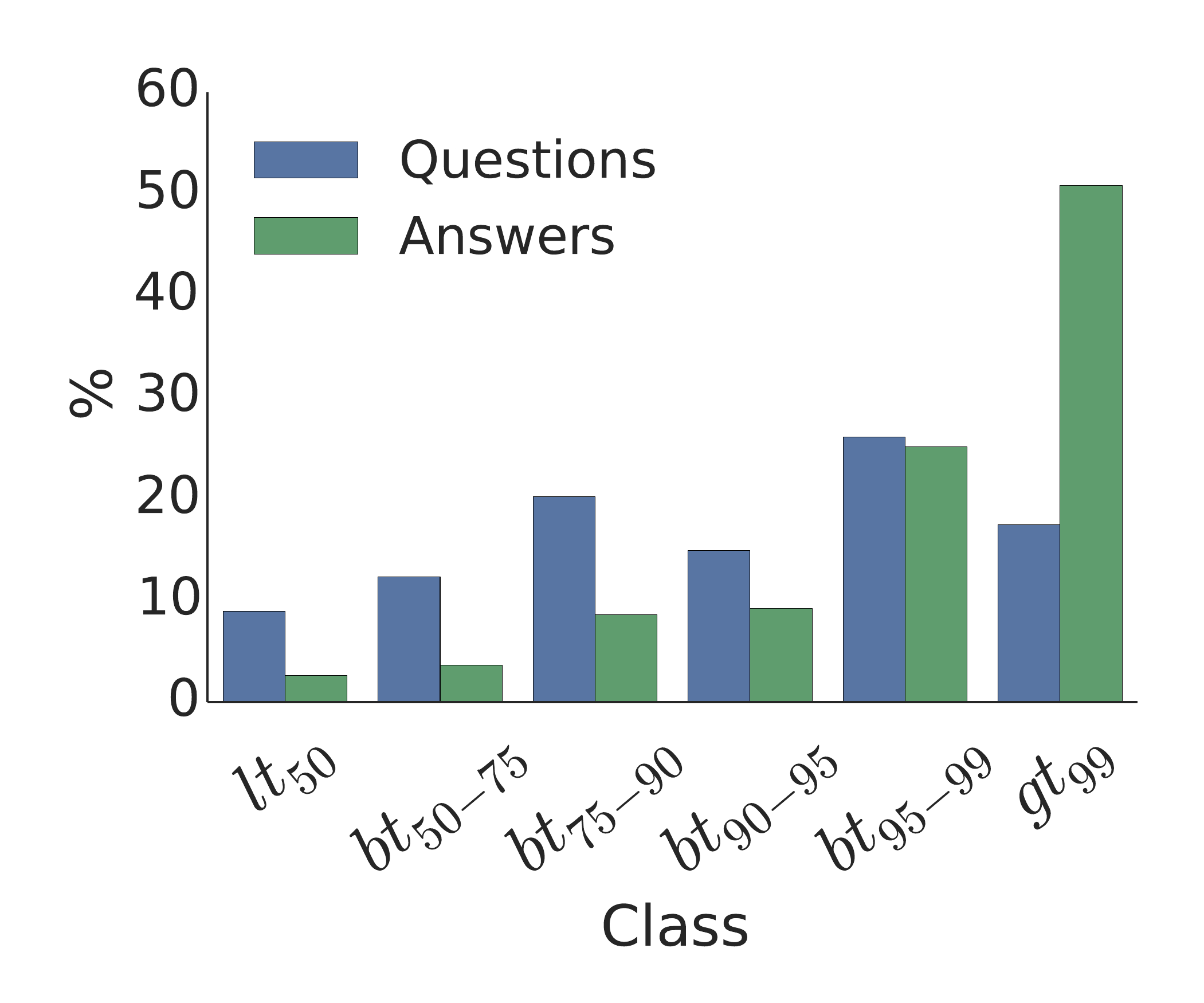}

  \end{minipage}
  \hfill
  \begin{minipage}[b]{0.5\textwidth}
        \vspace{3em}
    \centering
\scriptsize
\begin{tabular}{C{1cm}|L|L|C{1.3cm}}\hline
    Category & Questions & Answers & Total \\ \hline
        LC & $6.6 \times 10^6$ \textbf{(56.43\%)} & $4.5 \times 10^6$ \textbf{(24.08\%)}  & $1.11 \times 10^7$\\
    MC & $3.0 \times 10^6$ \textbf{(26.10\%)} & $4.7 \times 10^6$ \textbf{(25.09\%)} & $7.7\times 10^6$\\
    TC & $2.0 \times 10^6$ \textbf{(17.46\%)} & $9.6 \times 10^6$ \textbf{(50.51\%)} &$1.16\times10^7$\\
    \hline
\end{tabular}
      \vspace{6.7em}
    \end{minipage}
    \caption{(Left) Relative contribution of the users of percentile classes in questioning and answering. From $lt_{50}$ to $gt_{99}$, the contribution in answering increased, while in questioning, it seemed to decrease. (Right) Category-wise contribution across questioning and answering. Only a small proportion of questions was contributed by TC class.}
        \label{fig:qa_bars}
  \end{figure}
      \vspace{1em}
%
%

Since our classification of users is based on their total contribution in questioning and answering, i.e., their qa-count, we examined which of these two activities was contributed more by each class. We, therefore, computed the fraction of the total questions and answers posted by the users of these classes.
The bar plot in Figure~\ref{fig:qa_bars} shows that the lower classes' users are contributing more in questioning than answering. On the other hand, the contribution of users with the highest qa-count (i.e., $gt_{99}$ class) is more in answering. Further, aggregating the numbers for LC-MC-TC categorization,
the Table in Figure~\ref{fig:qa_bars} shows that 
although the contribution in answering by TC (50.51\%) is quite higher than by MC (25.09\%) and LC (24.08\%), the contribution in questioning by TC (17.46\%) is much less than MC (26.10\%) and LC (56.43\%). The last column in the Table also shows that although the total quantitative contribution by LC and TC is comparable, the former class is more into questioning, while the latter is inclined towards answering. This shows that highly active users are mostly providing answers, whereas, the questions are mostly coming from the users whose overall interaction with the portal is quite low. Further, the contribution of MC users in questioning is found to be more or less comparable to their answering contribution. 

\subsubsection{Qualitative Analysis}

%
%
%
%
%
%
%


The quality of questions and answers may be measured by the votes that they accumulate. An additional measure for examining the worth of a question is the number of users that mark the question as favorite. 
We, therefore, computed these quality measures for the contribution made by the users in the six classes, i.e., the votes on answers ($a_{votes}$), the votes on questions ($q_{votes}$) and the number of their questions that were marked as favorite ($q_{fav}$). Table~\ref{quality} shows the proportion of $a_{votes}$, $q_{votes}$ and $q_{fav}$ for each class as well as LC-MC-TC categories.
TC users were found to be outperforming in the total number of $a_{votes}$ with LC, MC and TC collecting a total of 17.37\%, 21.79\% and 60.80\% votes on their answers respectively. These values are also coupled with the fact that the fraction of answers provided by TC was also much higher as compared to LC users (refer to Subsection~\ref{quant}). Further, the computation of average number of votes on answers provided by the classes revealed a reasonably high value of average $a_{votes}$ by the class $lt_{50}$ (2.30 $\pm$ 15.25) which was not very low as compared to $gt_{99}$ (2.92 $\pm$ 20.23). This points towards the existence of a few subject experts casually indulging in providing one or two high quality answers (2 being the maximum qa-count of $lt_{50}$), increasing the average $a_{votes}$ of the class. On the other hand, LC users gathered slightly more $q_{votes}$ (34.83\%) than MC (31.88\%) and TC (33.26\%) as well as more $q_{fav}$ (35.07\%) than 
MC (31.63\%) and TC (33.27\%) respectively.  
Overall, this suggests that while LC users were providing a good proportion of quality questions, TC users are providing good quality answers. Further, MC seemed to be having a mixture of users engaged in questioning or answering or both, resulting in the values of $q_{votes}$, $q_{fav}$ and $a_{votes}$ that lie between those for LC and TC.

\begin{table}
\centering
\includegraphics[scale=0.5]{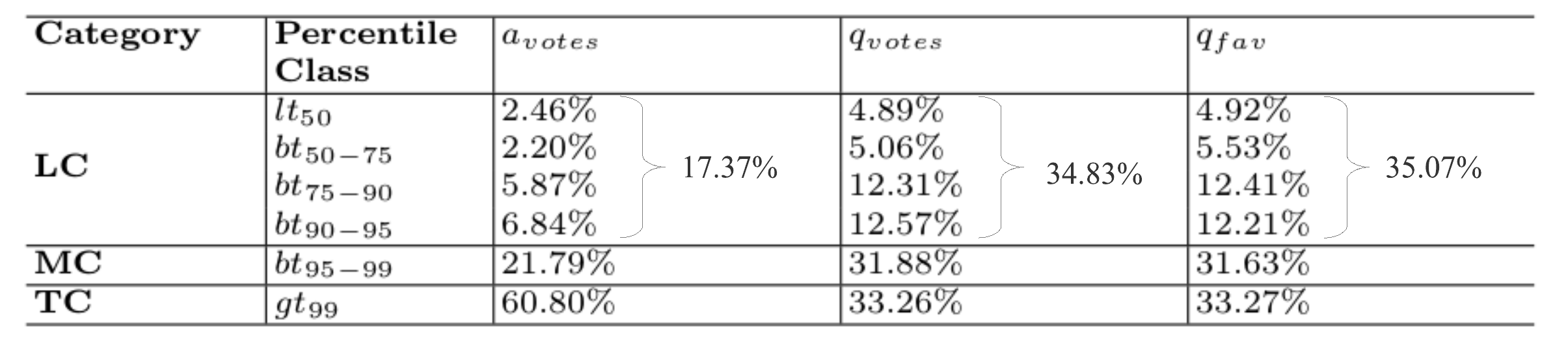}
\caption{Fraction of $a_{votes}$, $q_{votes}$ and $q_{fav}$ accumulated by the users of different classes}
\label{quality}
\end{table}

\subsubsection{Dependency Among Users from Different Categories}

\begin{figure}[h]
\begin{subfigure}{0.5\textwidth}
\includegraphics[scale=0.35]{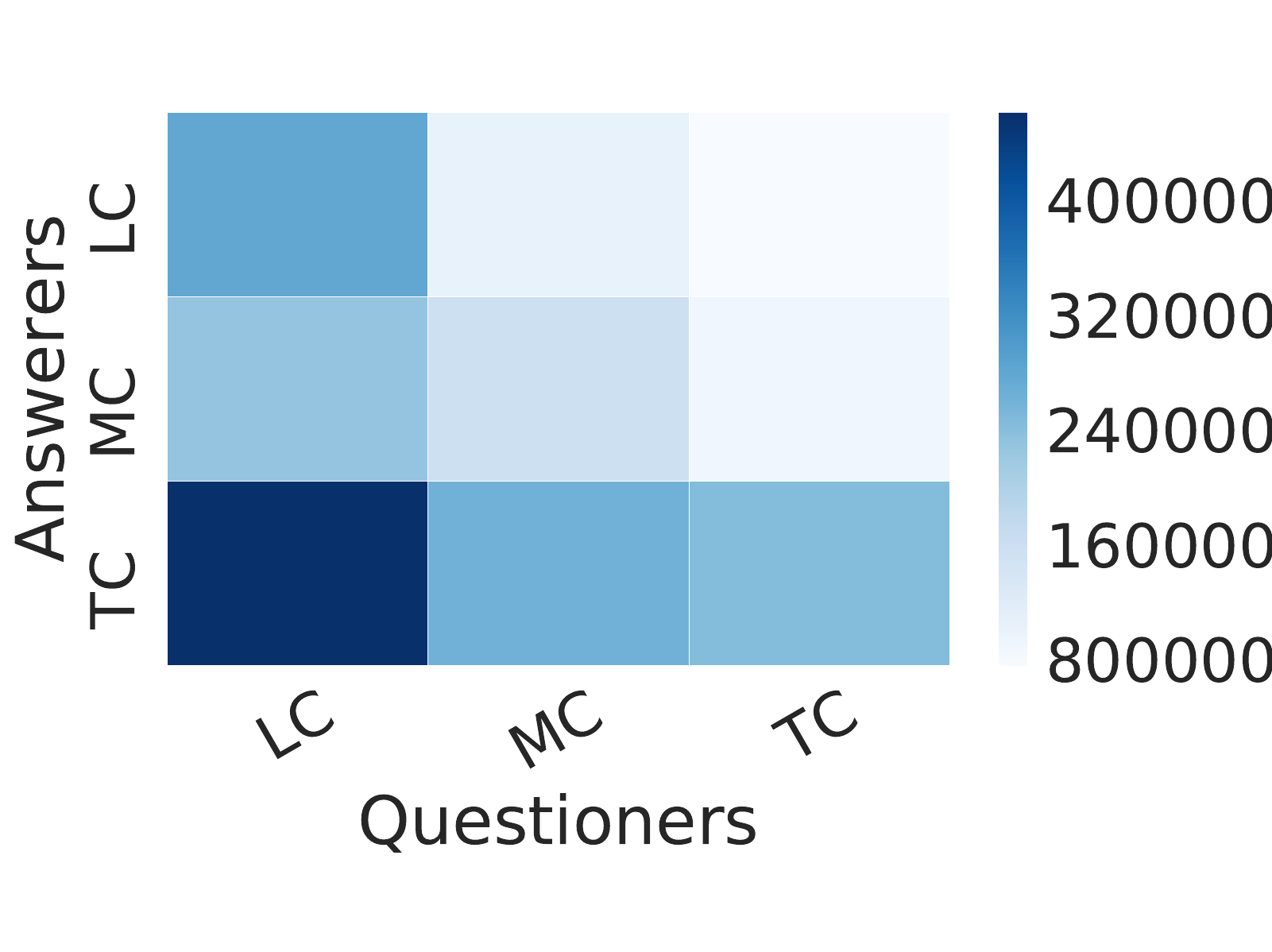}
\vspace{-2em}
\caption{}
\label{fig:heatmap_SO}
\end{subfigure} 
\begin{subfigure}{0.5\textwidth}
\includegraphics[scale=0.35]{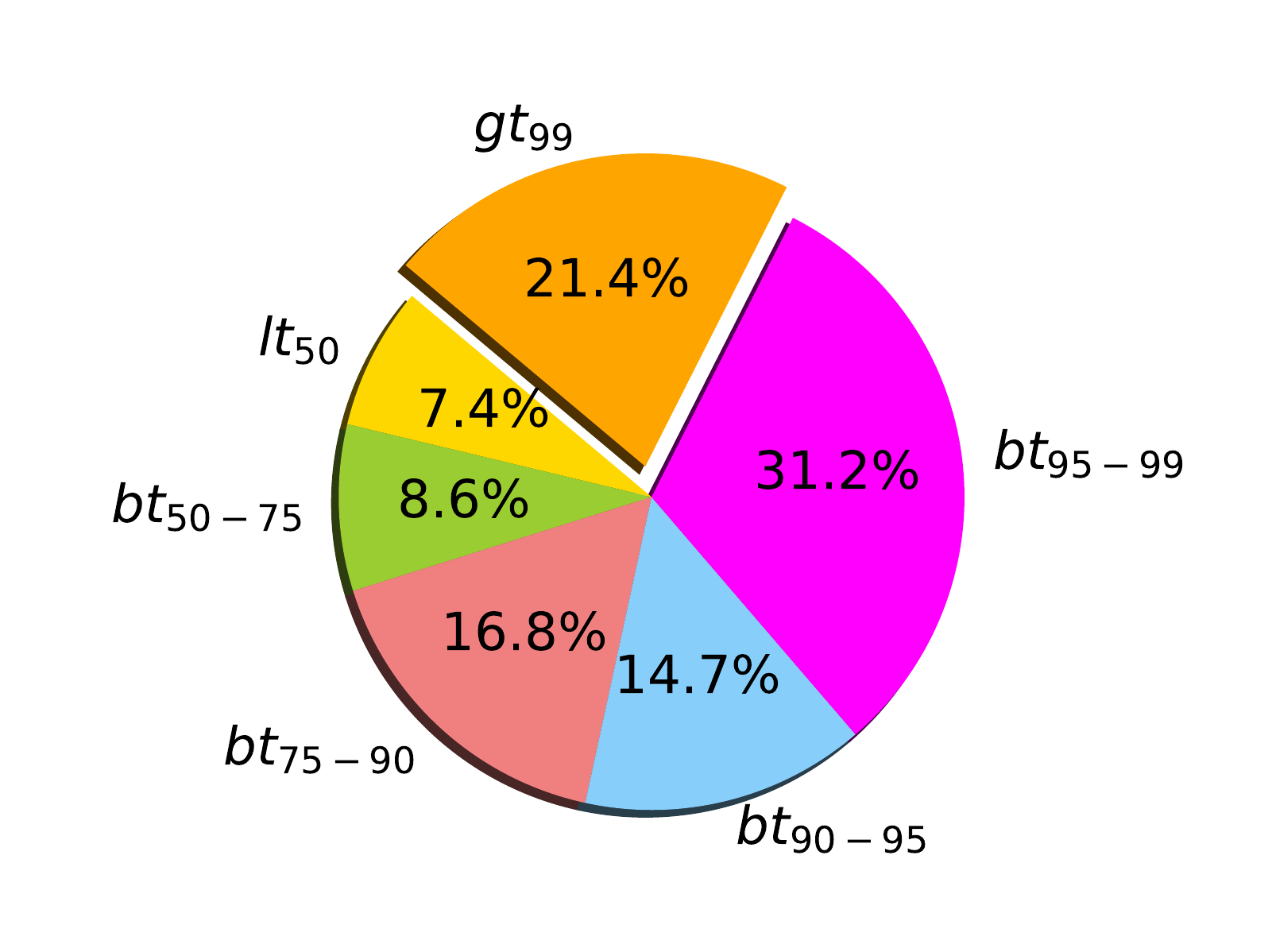}
\caption{}
\label{fig:pie}
\end{subfigure}
\caption{(a) 
Number of questions answered by the users of the category on the Y-axis that were asked by the users of the category on the X-axis. The bottom-left cell indicates high dependency of TC users on LC users for questions. (b) Fraction of questions from different percentile classes that were answered by the top contributor $\mathcal{T}$.}
\label{fig:both}
\end{figure}

We examined the dependency across the content added by the three categories by checking the users involved in the question-answer pairs. 
%
For each such pair, we noted the category of the user who asked the question (i.e., \textit{questioner}) and the user who answered it (i.e. \textit{answerer}). We examined a total of 19,090,945 \textit{questioner}-\textit{answerer} pairs. We first identified the category of each user of the pair and then computed the total number of questions answered by a given category where the questions were asked by the other category. We computed this for all $3\times3$ different combinations of categories. The results are shown in Figure~\ref{fig:heatmap_SO} where the labels on the X-axis indicate the category of the questioner and the labels on the Y-axis show the category of the answerer. The intensity of color in the cells represents the total number of questions that were asked by the users of the category on the X-axis that were answered by the users of the category on the Y-axis. The bottom most row, for example, depicts the number of questions answered by TC users that were asked by the users of other categories. It can be seen that 
out of all the questions answered by TC, a large part of the questions was coming from LC category users, while only a small portion of these questions was being asked by TC users. In particular, out of all the questions answered by TC, 47.54\% were asked by LC, 27.20\% were asked by MC whereas only 25.23\% were asked by TC.
This points to the dependency of TC users on the LC users who provide them a large proportion of the questions answered by them. 


\vspace{1cm}
\noindent \textbf{Case study of the top contributor: }
We already observed that the users with a large qa-count are mostly providing answers and asking fewer questions (Refer to Figure~\ref{fig:qa_bars}). In particular, the top contributor as per the qa-count, asked 39 questions and posted 33135 answers. Extending the preceding analysis of examining dependency, we inspected this top contributing user $\mathcal{T}$ as a representative case to explore its dependency on the users of other classes. We examined the qa-count of the users whose questions had been answered by $\mathcal{T}$ and hence identified their percentile class. Overall, $\mathcal{T}$ had answered questions of 23,892 users. 
Figure~\ref{fig:pie} shows the fraction of questions asked by each class that were answered by $\mathcal{T}$. Further, summing up the values category wise, only 21.4\% of these questions were asked by the TC users, while LC users asked a total of 47.5\% of these questions.
This case study highlights that top users providing answers are dependent to a good extent on the low contributing users for questions.


\subsection{HITS Analysis of the Q\&A Network of StackOverflow}\label{ha}

%
We now use another approach to examine the basic traits of the users in each class. This approach is based on the link analysis of the network formed out of the interaction among users. Link analysis is widely used for finding authoritative nodes in a network~\cite{bellomi2005network}. Particularly for a web graph, the most commonly used algorithms are PageRank~\cite{page1998pagerank}, HITS (\textit{Hyperlink-Induced Topic Selection})~\cite{kleinberg1999authoritative} and SALSA~\cite{lempel2000stochastic}. Out of these, HITS computes two values for each node, (i.e., a web page): \textit{Hub} score and \textit{authority} score. Here, authorities are web pages that contain useful information on a topic whereas hubs are web pages that point to useful authorities, i.e., web pages that are useful for a search query. The algorithm follows a recursive procedure which starts out by assigning a hub value of 1 to each node. The authority value of all the nodes is then computed as the sum of the hub values of the nodes that point to them. The hub value of each node is further updated by computing the sum of the authority values of the nodes that are pointed to by it. Formally, if $G = (V, E)$ is the underlying directed graph where $V$ is the set of hyperlinked web pages and $E$ is the set of directed edges among them, then each page $p$ has a hub score $p_h$ and an authority score $p_a$. The following two operations are then applied on these two values of all the nodes:

\begin{minipage}{0.5\textwidth}
\begin{equation*}
p_a = \sum_{q:(q,p)\in E}q_h
\end{equation*}
\end{minipage}
\begin{minipage}{0.5\textwidth}
\begin{equation*}
q_h = \sum_{p:(q,p)\in E}p_a
\end{equation*}
\end{minipage}

By means of these operations, the hub and authority scores are mutually reinforced. Further, during each iteration, the values are normalized. It has been proven that by applying the above operations in an alternating way, the hub and authority scores converge after a certain number of iterations. 

\begin{figure}[!htb]
\begin{minipage}[b]{0.4\textwidth}
\hspace*{-2cm}
\includegraphics[scale=0.3]{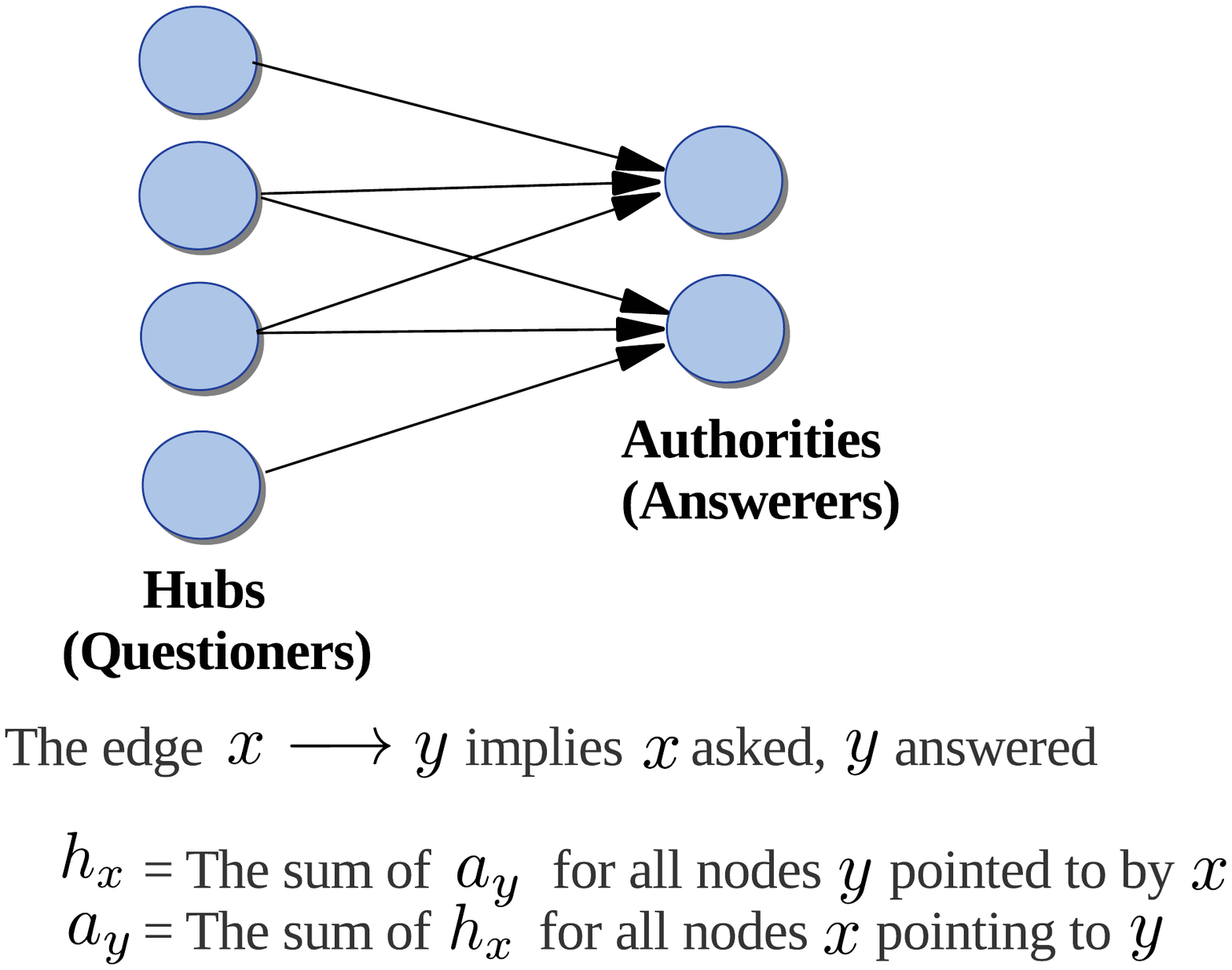}
\vspace*{-2cm}
\captionof{figure}{Description of edges and HITS analogy of a web graph for a Q\&A network.}
\label{fig:HITS-analogy}
\end{minipage}\hfill
\begin{minipage}[b]{0.55\textwidth}
\centering
\includegraphics[scale = 0.3]{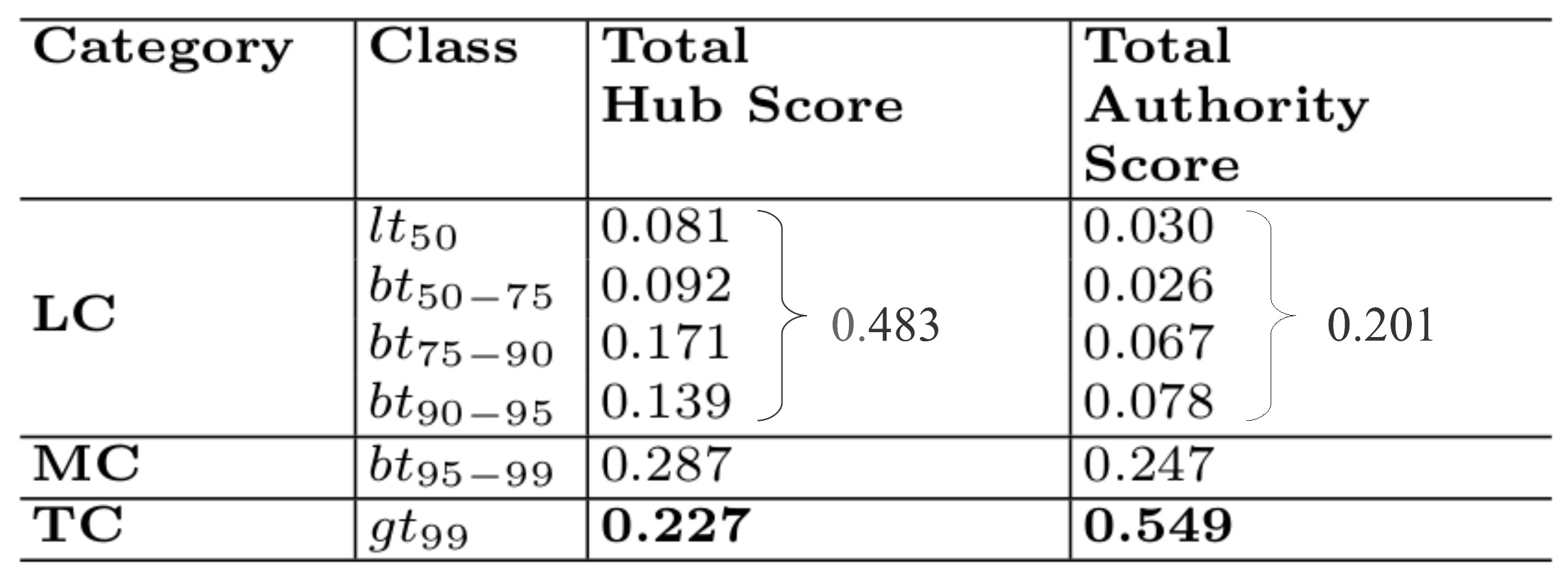}
\vspace*{0.8cm}
\captionof{table}{Total hub and authority scores of the classes. TC users were leading in authority score, while LC users had a higher hub score. \vspace*{0.15cm}}
\label{tab:hub_auth}
\end{minipage}
\end{figure}

Apart from web graphs, HITS approach finds an equivalent parallel in Q\&A networks where the nodes are users and a node $x$ points to another node $y$ if $x$'s question has been answered by $y$ (See Figure~\ref{fig:HITS-analogy}). In a web graph, hubs are web pages that point to good authorities and authorities are web pages that are pointed to by good hubs. Analogously, in a Q\&A network, \textit{questioners} are good if their questions are answered by good \textit{answerers}. Similarly, \textit{answerers} are good if they answer questions asked by good \textit{questioners}. Therefore the hub and authority scores directly map to the questioning and answering capabilities of users respectively. 
This approach has been used by Jurczyk et al.~\cite{jurczyk2007discovering} who used HITS for finding authoritative users in YahooAnswers, i.e., the users who are good in providing answers. Acknowledging the importance of questions in a Q\&A portal, we extend their approach and use HITS to find both the authority score (answering capability) as well as hub score (questioning capability) of the users.

Table~\ref{tab:hub_auth} shows the total
hub and authority scores obtained for each class.
As expected, the authority score of the top class was the highest. However, there were a few other observations as noted below:
\begin{enumerate}
\item The hub score of the top class (0.227) was quite less than its authority score (0.549), indicating that their questioning capabilities on the portal were not as good as answering. 
\item For the lower classes, hub score was more than the authority score, indicating that 
their questioning capability was better than their answering capability. In particular,
LC (0.483) was having a better hub score than MC (0.287) and TC (0.227), depicting a better questioning capability of users with less interaction with the portal as compared to the bunch of highly active experts.
\item As observed previously, MC users were found to be contributing in both questioning as well as answering shown by their comparable hub (0.287) and authority scores (0.247). 

\end{enumerate}

The hubs and authorities analysis, investigating the network structure of interaction among users, substantiates the qualitative and quantitative analysis conducted in the previous Subsections. It confirms the importance of a large mass of users interacting with the system less often. They provide a useful input to Q\&A systems in the form of questions, which are then answered by the elite users, leading to the creation of a useful knowledge repository. 
The masses thus appear to be a useful part of the Q\&A ecosystem. 

\section{Discussion}


The analysis examines the contribution made by different strata of users in Q\&A portals in an attempt to investigate the debatable Ortega Hypothesis. In contrast with the latest work on Ortega Hypothesis which was conducted in the scientific domain~\cite{bornmann2010scientific}, we observe that in Q\&A environments, the users making less quantitative contribution are \textit{also} required in the system. This is because they are found to be a source of a substantial proportion of questions. Our analysis on StackOverflow highlighted the dependency of expert answerers on masses for their questions. Due to their innate characteristics and a good domain knowledge, experts who actively provide answers are less likely to ask questions. It is the large number of infrequent users who visit the website with lesser frequency to get their problems solved, are the main source of questions. These questions help in extracting knowledge from the expert users in the elite bunch, hence in building a comprehensive knowledge-base. Further, it is understandable that it is not possible for a small bunch of expert users to come up with a diverse set of questions on problems that the less-experienced users might come across. Such questions are more likely to come from a large group of users who are not specially-qualified (i.e., masses). The hub and authority scores of the classes further substantiated the lack of questioning capability of elite users. Therefore, particularly in the context of Q\&A portals, the masses seem to hold an important role. 

The study shows that the elite users due to their higher answering capability seem to be the experts in the topic of the portal. If a group is formed out of only such experts in a Q\&A portal (as per Newton hypothesis), it will be difficult to build knowledge due to a lack of possible questions. 
Therefore, our results incline towards refutation of Newton hypothesis in Q\&A portals that supports the idea of just a bunch of elites being able to run the system. In other words,
if Q\&A portals are built following the claims of Newton Hypothesis
such systems might fail to achieve their intended purpose.
Further, the results support the claims of Ortega hypothesis, observing the presence of masses to be important for Q\&A systems. At the same time, the study also finds elite users to be responsible for providing answers to the largest proportion of the questions asked. This indicates that the two hypotheses may not be mutually exclusive and that, a combination of these may be true in practice. 



It is important to remark that it may be possible that the elite users identified in the analysis behave differently in the context of a different setup. Therefore, the results of the current study as well as the conclusions made are restricted to Q\&A portals. It is also important to remark that despite a disparate difficulty level associated with questions and answers, we represent them equally. This is a deliberate choice made in order to focus on examining the basic traits of users across the spectrum. The analysis may be extended by tracking users' contribution in other activities available on the portal such as commenting and voting to obtain a higher dimensional perspective. Further, it will be useful to note that the results are bound to the percentile classes we used to obtain the stratification of users. Although this approach seems reasonable enough, the possibilities of any alternate classification providing additional insights may not be discounted.

The study highlights that the online communities are not egalitarian and hence
suggests tapping the potential of users through stratified mechanisms. The steps could be taken both at the interface level as well as at the level of devising
rewarding mechanisms. 
For instance, the interface design should be such that it eases usage for users visiting less often as well as facilitates advanced tools and features for more active and dedicated users.
Incentivization strategies that reward users' achievements based on a system of badges or points are known to affect users' participation level~\cite{anderson2013steering,cavusoglu2015can}. 
Therefore, these policies need to be carefully drafted, as inappropriate incentivization policies may,
in turn, hamper the system's functioning~\cite{yang2014sparrows,merton1968matthew}. The existing policies have largely been built acknowledging only the contribution by elites, i.e. generally considering Newton Hypothesis to be true. Further, such policies are likely to lead to a \textit{rich-get-richer} phenomenon~\cite{merton1968matthew} 
which may result in an inefficient utilization of potential contributors making the `Wisdom of crowds'~\cite{surowiecki2005wisdom} effect fade away. 
To handle this, the progress monitoring mechanisms may bracket the users in different strata as per their contribution and may incentivize them differently. This can be achieved through milestones with gradually scaling difficulty rather than a uniform and an evenly-spread reward system. A relatively easily achievable reward system for beginners will encourage them to contribute more.
Another guideline requiring both interface and policy level changes involves highlighting not only the top users on the interface but also those that are doing well out of the low contributing cohort. Such a space on the interface dedicated for pacing up newcomers may help in retaining them on the portal.

\section{Conclusion}
The work investigates the worth of masses in Q\&A portals, where masses are users whose interaction with the portal is very less as compared to a bunch of highly active elites. Our study shows that masses provide useful contribution to the system and are also responsible for extracting more contribution from elites thus supporting the claims of Ortega hypothesis in Q\&A portals. The results also suggest that a Q\&A system, whose policies are built considering Newton hypothesis to be valid, i.e., keeping only the bunch of elite users in mind, will not be able to produce optimal results. The study, therefore, recommends examining and redefining system policies for tapping the potential of masses.






\bibliographystyle{unsrt}  
\bibliography{/home/anamika/Dropbox/0_PhD/BIB_PATH/KB_bibliography}

\end{document}